\documentclass[12pt]{article}

\usepackage[totalheight = 23cm, totalwidth = 17cm]{geometry}
\usepackage{amssymb,amsmath,amsfonts,amsbsy,cite}
\usepackage{bm}
\usepackage{color}
\usepackage{subfigure}

\usepackage{color}
\usepackage{graphicx}
\usepackage{calc}

\numberwithin{equation}{section}
\newcommand{\mpl}{M_{\text{Pl}}}

\newcommand{\be}{\begin{equation}}
\newcommand{\ee}{\end{equation}}
\newcommand{\bea}{\begin{eqnarray}}
\newcommand{\eea}{\end{eqnarray}}
\newcommand{\nn}{\nonumber}

\begin{document}
\begin{titlepage}

\begin{flushright}
    YITP-17-135
\end{flushright}
\begin{center}
\vskip 1cm
\LARGE{\bf Scalaron from $R^2$-gravity as a Heavy Field}
\vskip 1cm
\large{Shi Pi$^{a}$,~Ying-li Zhang$^{b}$,~Qing-Guo Huang$^{c,d,e,f}$,~and~Misao Sasaki$^{a,g}$}

\vskip 1.2cm

\small{\it
$^{a}$ Center for Gravitational Physics, Yukawa Institute for Theoretical Physics, Kyoto University, \\
Kyoto 606-8502, Japan
\\
$^{b}$ Department of Physics, Faculty of Science, Tokyo University of Science, \\
1-3, Kagurazaka, Shinjuku-ku, Tokyo 162-8601, Japan
\\
$^{c}$ CAS Key Laboratory of Theoretical Physics, Institute of Theoretical Physics, Chinese Academy of Sciences, \\
55 Zhong Guan Cun East Street, 100190, Beijing, China
\\
$^{d}$ School of Physical Sciences, University of Chinese Academy of Sciences, \\
No. 19A Yuquan Road, Beijing 100049, China
\\
$^{e}$ Synergetic Innovation Center for Quantum Effects and Applications, Hunan Normal University, \\
36 Lushan Lu, 410081, Changsha, China
\\
$^{f}$ Center for Gravitation and cosmology, College of Physical Science and Technology, Yangzhou University, \\
88 South University Ave., 225009, Yangzhou, China
\\
$^{g}$ International Research Unit of Advanced Future Studies, Kyoto University, Kyoto 606-8502, Japan\\
\vspace{2em}
}

\vskip 1.2cm

\end{center}

\begin{abstract}
We study a model of inflation in which a scalar field $\chi$ 
is non-minimally coupled to Starobinsky's $R^2$ gravity. 
After transforming it to the Einstein frame, 
a new scalar field, the scalaron $\phi$, will appear and couple to $\chi$ 
with a nontrivial field metric, while $\chi$ acquires a positive mass 
via the non-minimal coupling. Initially inflation occurs along the $\phi$ direction 
with $\chi$ trapped near its origin by this induced mass. After $\phi$ crosses 
a critical value, it starts rolling down rapidly and proceeds to
damped oscillations around an effective local minimum determined by 
the value of $\chi$, while inflation still continues, driven by the $\chi$ field 
at this second stage where the effect of the non-minimal coupling becomes 
negligible. The presence of the damped oscillations during the
transition from the first to second stage of inflation
causes enhancement and oscillation features in the power spectrum
of the curvature perturbation. 
Assuming that the oscillations may be treated perturbatively,
we calculate these features by using the $\delta N$ formalism,
and discuss its observational implications to
large scale CMB anomalies or primordial black hole formation,
depending on the scale of the features.

\end{abstract}

\end{titlepage}

\setcounter{page}{0}
\newpage
\setcounter{page}{1}

\section{introduction}
Inflation is a stage in the early universe during which it has expanded almost
exponentially by a number of $e$-folds $N\gtrsim 60$,
which was originally meant to solve
several initial condition problems of the big-bang cosmology such as the initial
singularity, horizon and flatness problems
~\cite{Brout:1977ix,Sato:1980yn,Guth:1980zm,Starobinsky:1980te,inflation}. 
However, it was realized that the most important testable
consequence of inflation is the production
of scalar-type and tensor-type cosmological perturbations
from quantum vacuum fluctuations during 
inflation~\cite{Mukhanov:1981xt,Starobinsky:1979ty}.
During inflation, the Hubble horizon $H^{-1}$ is nearly a constant, 
while the quantum fluctuations generated on small scales are stretched 
out to superhorizon scales and become frozen, which then re-enter 
the horizon after inflation and seed the cosmic microwave background (CMB) 
anisotropy and the large scale structure  (LSS) we observe today. 

In the past two decades, observations have been improved  to such 
a high accuracy that the observed data not only confirm the predictions 
of inflation but also can be used to distinguish different inflation models. 
For instance, Planck 2015 has not detected any obvious signal of primordial 
tensor fluctuations, while the primordial scalar spectrum is red-tilted, 
with the scale-invariant Harrison-Zeldovich spectrum already excluded 
at $5.6\sigma$ confidence level~\cite{Ade:2015lrj}, which consequently
excluded the $\phi^2$ chaotic inflation model. 
According to the analysis of the inflation models, the concave potential 
is favored~\cite{Huang:2015cke}. Especially, Starobinsky's 
$R^2$-inflation \cite{Starobinsky:1980te} predicts the 
tensor-to-scalar ratio $r\sim 1/N^{2}$ and the spectral tilt $n_s\sim(1-1/N)$,
and sits right near the sweet spot of the Planck $r$-$n_s$ contour~\cite{Ade:2015tva}.

Recently the possibility of detecting the effects of
new particles during inflation has 
attracted much attention. Although the quasi-exponential 
expansion may be driven by a single scalar field, the inflaton,
it is plausible that there are other fields that play some secondary 
but non-trivial roles during inflation.
In string theory or supersymmetry/supergravity such scalar fields
seem to be ubiquitous during inflation. 
Light scalar fields ($m\ll H$) may modify the inflationary 
trajectory in a non-trivial way~\cite{twolightfield}. If they are heavy, 
the inflationary trajectory will not change at leading order,
but they can leave detectable signals on the power spectrum/bispectrum/trispectrum/etc.
To the leading order, the heavy fields can be integrated out~\cite{Tolley:2009fg,EFT}, which gives us an effective single field with a non-trivial speed of sound, like $k$-inflation~\cite{kinf}. Besides, ``nonlocal'' effects will produce some
 oscillatory features on the power spectrum/bispectrum~\cite{features,standardclock}, 
which can be viewed as imprints of heavy fields.
In these models, the inflaton usually rolls down
from a high plateau to a valley, and begins to oscillate around the 
bottom of the valley as it slowly moves along the valley.
These oscillations correspond to the excitations of heavy fields. 

In most cases the potential for a heavy field is 
assumed to be quadratic, $(1/2)M^2\phi^2$. A heavy field with a general 
$\phi^p$-potential has been studied in \cite{Huang:2016quc}. An interesting 
extension is to consider the scalaron of the 
Starobinsky model with a potential $(M/2)^2(1-e^{-\phi/\mpl})^2$
as a heavy field.
The Starobinsky model can originate from higher order curvature correctons
to the Einstein-Hilbert action at high energy scales~\cite{conformal}.
 This has motivated us to study the physics of scalar fields in $R^2$-gravity. 

A scalar field $\chi$ in $R^2$-gravity will have a non-trivial kinetic term 
coupled to the scalaron $\phi$~\cite{Starobinsky:2001xq}
when transformed to the Einstein frame. 
If we set its potential to be flat, inflation will firstly occur from 
the $R^2$-gravity induced scalaron potential plateau, and then 
the scalaron rolls down to its effective local minimum
determined by $\chi$, say $\phi=\phi(\chi)$, where $\phi$ becomes sufficiently 
heavy. Inflation continues as $\chi$ rolls down its effective potential
$V_{\rm eff}(\chi)=V(\chi,\phi(\chi))$. 
Such kind of models have been studied by analytical or 
numerical methods in for 
instance \cite{DiMarco:2002eb,Cardenas:2003tg,Tsujikawa:2004my,DiMarco:2005nq,Choi:2007su,vandeBruck:2014ata,Kaneda:2015jma,Wang:2016ipp,Mori:2017caa}, and
it has been shown that the scalar field with non-canonical kinetic term 
may generate interesting feature signals. 

In this paper we consider the case where the scalaron undergoes
damped oscillations after it has become heavy. 
The oscillations around the leading order trajectory are treated as 
a classical perturbation, and the corrections to 
the power spectrum of the curvature perturbation are derived analytically. 
Whether the transition from the $\phi$-domination to 
the $\chi$-domination occurs early or late during inflation
gives us different observational outcomes. 
If the transition happens right before the current horizon scale 
leaves the horizon, the oscillations from the transition would generate
 features on the large angular scale CMB anisotropy. On the other hand,
if the transition occurs late, i.e. if the scale that left the horizon at 
the time of transition is far below the CMB and LSS scales, 
we do not have any stringent observational constraints on the transition
and the subsequent stage of inflation. In this case,
an interesting possibility is the generation of primordial 
black holes (PBHs) when the modes which exit the horizon reenter. In our mode there will be a near monochromatic PBH mass function, which may account for the cold dark matter of the current Universe.

This paper is organized as follows. 
In Section \ref{sec:setup} we introduce our model based on the 
considerations above. We then calculate the background evolution 
and clarify the possible range of the model parameters. 
In Section \ref{sec:pert} we calculate the power spectrum of the 
curvature perturbation by using the $\delta N$ formalism.
We then discuss the possible 
observational impacts on the CMB anisotropy and PBH production in 
Section \ref{sec:cmb} and Section \ref{sec:pbh}, respectively. 
In Appendix  \ref{app:efold} , we give useful relations between the Hubble parameter $H$
at a couple of epochs during inflation and the Hubble constant today, $H_0$. In  Appendix \ref{app:pbh}, we derive a formula for the PBH mass in terms of relevant $e$-folding numbers.

\section{Set up and the Background Evolution}\label{sec:setup}

We consider the action of Starobinsky's $R^2$ gravity with 
a non-minimally coupled scalar in the Jordan frame,
\be\label{SJ}
S_J=\int d^4x\sqrt{-g}\left\{\frac{\mpl^2}{2}\left(R+\frac {R^2}{6M^2}\right)
-\frac12g^{\mu\nu}\partial_\mu\chi\partial_\nu\chi-V(\chi)-\frac12\xi R\chi^2\right\}.
\ee
This is a special case of $f(R)$ gravity with
\be
f(R)=R+\frac{R^2}{6M^2}-\xi R\frac{\chi^2}{\mpl^2}\,,
\quad
F(R)=\frac{\partial f}{\partial R}.
\ee
We will follow \cite{DeFelice:2010aj}, and define a new field 
$\phi=\mpl\sqrt{3/2}\ln F$, while the conformal transformation 
$\tilde g_{\mu\nu}=Fg_{\mu\nu}$ can transfer the action \eqref{SJ} to 
the one in Einstein frame, 
\bea\nonumber
S_E&=&\int d^4x\sqrt{-\tilde g}\cdot\left\{\frac{\mpl^2}2\tilde R-\frac12\tilde g^{\mu\nu}\partial_\mu\phi\partial_\nu\phi-\frac12e^{-\sqrt{\frac23}\frac\phi\mpl}\tilde{g}^{\mu\nu}\partial_\mu\chi\partial_\nu\chi\right.\\\nonumber
&&\quad\quad\quad\quad\quad\quad-\left.\frac34M^2\mpl^2\left(1-e^{-\sqrt{\frac23}\frac\phi\mpl}+e^{-\sqrt{\frac23}\frac\phi\mpl}\xi\frac{\chi^2}{\mpl^2}\right)^2-e^{-2\sqrt{\frac23}\frac\phi\mpl}V(\chi)\right\}.\\\label{SE}
\eea
From now on we will work only in the Einstein frame, and assume the 
Friedmann-Lemaitre-Robertson-Walker metric 
$\tilde g_{\mu\nu}=\text{diag}(-1,a^2,a^2,a^2)$, where $a(t)$ is the scale factor. 
We take the variation of \eqref{SE} with respect 
to $\tilde g_{\mu\nu}$, $\phi$ and $\chi$, and obtain the 
Friedmann equation as well as the equations of motion for $\phi$
 and $\chi$~\cite{Starobinsky:2001xq}:
\bea\label{Friedmann}
&&3\mpl^2H^2=\frac12\dot\phi^2+F^{-1}\frac12\dot\chi^2+\frac34M^2\mpl^2\left[1-F^{-1}\left(1-\xi\frac{\chi^2}{\mpl^2}\right)\right]^2+F^{-2}V(\chi),\\\nonumber
&&\ddot\phi+3H\dot\phi+\sqrt{\frac32}M^2\mpl F^{-1}\left\{1-\xi\frac{\chi^2}{\mpl^2}+\frac{\dot\chi^2}{3M^2\mpl^2}-F^{-1}\left[\left(1-\xi\frac{\chi^2}{\mpl^2}\right)^2+\frac{4V}{3M^2\mpl^2}\right]\right\}=0,\\\label{eom:phi}\\\label{eom:chi}
&&\ddot\chi+\left(3H-\sqrt{\frac23}\frac{\dot\phi}\mpl\right)\dot\chi+3M^2\left[1-F^{-1}\left(1-\xi\frac{\chi^2}{\mpl^2}\right)\right]\xi\chi+F^{-1}V'(\chi)=0,
\eea
where $H\equiv\dot a/a$ is the Hubble parameter, and $F=e^{\sqrt{2/3}\phi/\mpl}$ is 
the conformal factor we use to define the scalaron $\phi$. During the first stage 
of inflation, $\phi$ is large, so because of the suppression of the $F^{-1}$ factor, 
the energy density is dominated by the potential of $\phi$, which is nearly a 
constant $(3/4)M^2\mpl^2$. We suppose $\chi$ can also drive inflation when $\phi$ 
has been trapped in its VEV. So we also set the potential of $\chi$ be concave. 
The simplest version of a concave potential may be the upside-down parabola,
\be\label{V(chi)}
V(\chi)=V_0-\frac12m^2\chi^2+\cdots,
\ee
where the dots represent the terms in higher powers of $\chi$
that make $V(\chi)$ bounded from below, and terminate inflation at some 
non-zero vacuum expectation value $\chi_\text{v}$. We do not need them here 
since during the inflationary era, we mainly focus on the phenomenon of the regime near $\chi=0$. 
Hence, higher-order terms are irrelevant.

At the first stage, the potential of $\phi$ dominates the energy density, thus drives
 inflation. And the energy density contributed by the $\chi$-part is 
nearly a constant. The Hubble parameter, after neglecting the kinetic terms
 for $\chi$, is
\bea\nonumber
H^2&\approx&\frac{M^2}4\left[\left(1-e^{-\sqrt{\frac23}\frac\phi\mpl}\right)^2+e^{-2\sqrt{\frac23}\frac\phi\mpl}\frac4{\mu^2}\right.\\\label{H>}
&&+\left.\left(2\xi\left(e^{-\sqrt{\frac23}\frac\phi\mpl}-e^{-2\sqrt{\frac23}\frac\phi\mpl}\right)-e^{-2\sqrt{\frac23}\frac\phi\mpl}\frac{2m^2}{3M^2}\right)\left(\frac\chi\mpl\right)^2\right].
\eea
where $\mu$ is defined as the mass $M$ divided by the Hubble parameter
 when $\phi\rightarrow0$,
\be\label{def:mu}
\mu^2\equiv\frac{3M^2\mpl^2}{V_0},
\ee
which is a key parameter in our model.
The Hubble parameter is just $M/2$ when $\phi$ is large ($F\gg1$) in the first stage of inflation driven by $\phi$. For our model 
of small-field inflation at the second stage, we have $\chi/\mpl\ll1$, which 
as well as $\xi\ll1$ and $m\ll M$ guarantees that the second line in \eqref{H>} 
is negligible compared to the first line. This means $H$ depends solely on $\phi$ 
at leading order during the first stage.

\begin{figure*}
  \centering
  \includegraphics[width=.5\linewidth - 0.25mm]{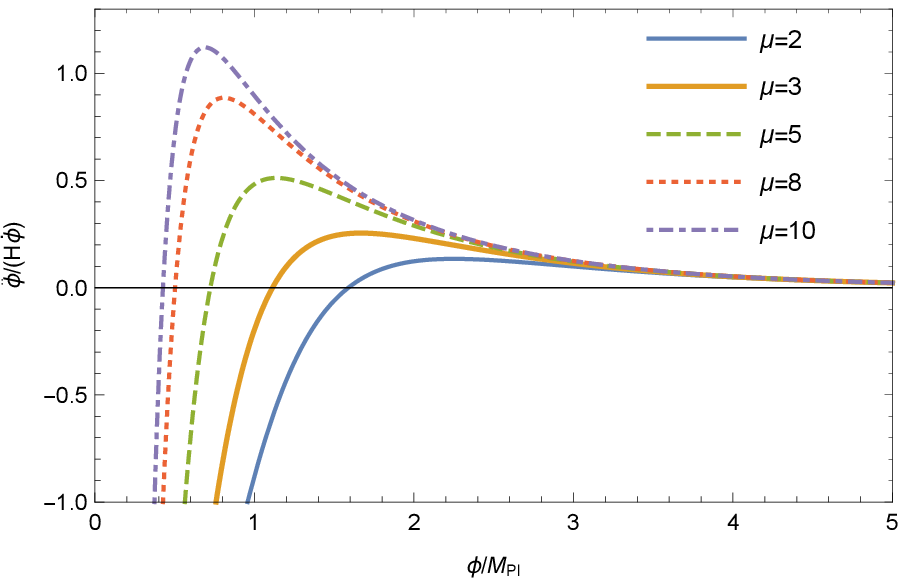}\hfill
  \includegraphics[width=.5\linewidth - 0.25mm]{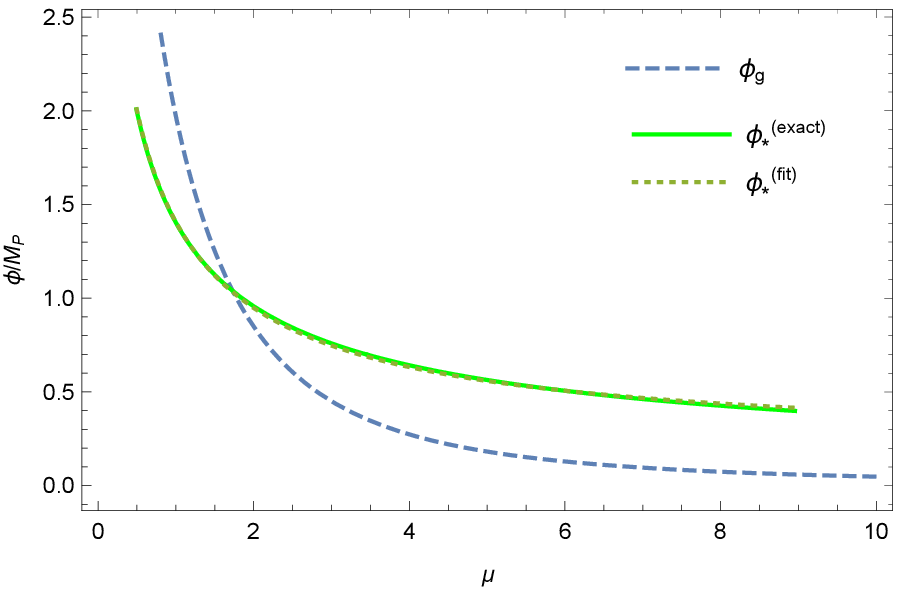}
  \caption{Left: The dependence of $\phi$ for $\frac{\ddot\phi}{H\dot\phi}$ at the first stage for different $\mu$ values. We see that for small $\mu$, slow-roll condition of $\phi$ breaks down at $\ddot\phi/(H\dot\phi)=-1$. Right: The critical point $\phi_*$ when $\frac{\ddot\phi}{H\dot\phi}=1$. The green curve is the exact solution while the dotted curve is the fitting \eqref{F*fit}. The dashed line dummed $\phi_g$ is the effective vacuum of $\phi$ at the second stage where it will be attracted to after the slow-roll condition is violated, thus the displacement between $\phi_*$ and $\phi_g$ is the initial amplitude for the oscillations.}\label{fig:phi}
\end{figure*}

Based on the same estimation and under the slow-roll condition, 
the equations of motion (\ref{eom:chi}) and (\ref{eom:chi}) reduce to the following forms, respectively
\bea\label{phisl}
&&3H\dot\phi=-\sqrt{\frac32}e^{-\sqrt{\frac23}\frac\phi\mpl}\left(1-e^{-\sqrt{\frac23}\frac\phi\mpl}\left(1+\frac4{\mu^2}\right)\right)M^2\mpl,
\\\label{chisl}&&\left(3H-\sqrt{\frac23}\frac{\dot\phi}{\mpl}\right)\dot\chi+3M^2e^{-\sqrt{\frac23}\frac\phi\mpl}\left[\xi \left(e^{\sqrt{\frac23}\frac\phi\mpl}-1\right)-\frac{m^2}{3M^2}\right]\chi=0.
\eea
Here we see that there are no $\phi$-$\chi$ coupling terms in \eqref{phisl} 
as they are controlled by $\mathcal O(\xi)$. This simplifies the equations 
and makes $\phi$ only depend on itself, which then sources $\chi$. 
We define the $e$-folding number,
\be\label{def:N}
N=\int^{t_*}_t Hdt\,,
\ee
as the new time variable. Note that it is counted backwards in time, 
so that $dN=-Hdt$. The upper limit for the integral $t_*$ is fixed at the end of 
the first stage, which is defined as the epoch when the $\phi$ field stops slow 
rolling, $|\ddot\phi/(H\phi)|_{t_*}=1$. Therefore, after changing the time 
variable in \eqref{phisl} and \eqref{chisl} to $N$, we have
\bea\label{phisl(N)}
\frac1\mpl\frac{d\phi}{dN}&=&2\sqrt{\frac23}\frac{F^{-1}-F^{-2}\left(1+\frac4{\mu^2}\right)}{1-2F^{-1}+F^{-2}\left(1+\frac4{\mu^2}\right)},\\\label{chisl(N)}
\frac1\mpl\frac{d\chi}{dN}&=&\frac{4\left[\xi(1-F^{-1})-F^{-1}\frac{m^2}{3M^2}\right]
}{1-\frac{14}9F^{-1}+\frac59F^{-2}\left(1+\frac4{\mu^2}\right)}\frac\chi\mpl.
\eea
We note that \eqref{phisl(N)} only depends on $\phi$, 
which can be easily integrated, 
\be\label{phihidden}
e^{\sqrt{\frac23}\frac\phi\mpl}-e^{\sqrt{\frac23}\frac{\phi_*}\mpl}-\sqrt{\frac23}\frac{\phi-\phi_*}\mpl+\frac4{\mu^2}\ln\frac{e^{\sqrt{\frac23}\frac\phi\mpl}-1-4/\mu^2}{e^{\sqrt{\frac23}\frac{\phi_*}\mpl}-1-4/\mu^2}=\frac43(N-N_*).
\ee
The solution $\phi(N)$ is the inverse function of the relation above, 
which can be approximated as
\be\label{phiapprox}
\phi=\mpl\sqrt{\frac32}\ln\left(\frac43(N-N_*)+F_*\right),
\ee
when $\phi$ is large. Here the suffix $_*$ denotes the epoch when the first slow-roll stage
ends and the oscillatory stage begins, at which $|\ddot\phi/(H\phi)|=1$. 
We have
\begin{equation}
\frac{\ddot\phi}{H\dot\phi}=-\frac{d^2\phi/dN^2}{d\phi/dN}-\frac{d}{dN}\ln H=\frac4{3}\cdot\frac{(F-1)^3\mu^4-4(2F^2-3F+2)\mu^2-16}{\left[(F-1)^2 \mu ^2+4\right]^2}.\label{eta<}
\end{equation}
We show this ratio in Fig.~\ref{fig:phi}, where for 
$\mu\lesssim8.955$, $\ddot\phi/(H\dot\phi)$ never exceeds 1 and 
the slow-roll condition for $\phi$ is violated later at $\ddot\phi/(H\dot\phi)=-1$, 
after which $\phi$ starts oscillating. This is because when $\mu$ is small, 
the effective potential in the large $\phi$ region is not steep enough 
to violate the slow-roll condition before $\phi$ starts to oscillate. 
We will focus on this range of $\mu$ for simplicity. Note 
that, at this point $\phi_*$ is still larger than the effective vacuum value, 
where $\ddot\phi+3H\dot\phi\simeq0$. Set $\ddot\phi/(H\dot\phi)=-1$, we can see that $F_*(\mu)$ is the solution of an algebraic equation, whose solution can be approximated by
\be\label{F*fit}
F_*(\mu)\simeq1.184+\frac{1.969}\mu.
\ee
The accuracy in the parameter space we are interested in is 
almost perfect as is shown in Fig.~\ref{fig:phi}.
This relation \eqref{F*fit}, together with its derivative 
\eqref{phisl(N)} at $\phi_*$, 
\be\label{phidot*}
\left.\frac1\mpl\frac{d\phi}{dN}\right|_{\phi=\phi_*}
=\frac{4}{3}\sqrt{\frac32}\frac{F_*-1-\frac4{\mu^2}}{F_*^2-2F_*+1+\frac4{\mu^2}}
\ee
are the initial conditions for the evolution of $\phi$ in the second stage.

Now we turn to discuss the motion of $\chi$. We can see the non-minimal
coupling to $\chi$ keeps the initial condition of $\chi$ at the beginning 
of the second stage  be small. From \eqref{chisl(N)}, we know that 
the right hand side is proportional to $\xi(F-1)-m^2/(3M^2)$ if we neglect 
the smaller $\chi^3$-term. 
Then for large $N$,~$F\gg1$, and the r.h.s. of \eqref{chisl(N)} can be 
estimated by $\xi F$, and thus positive, while for small $N$,~$F\simeq1$, 
it is left with $-m^2/(3M^2)$ and negative. This means that $\chi$ decreases 
in the early stage, while it increases at later time with a small initial 
condition guaranteed by its evolution earlier. The critical value for 
$\chi$ to start increasing can be achieved by requiring the r.h.s. of 
\eqref{chisl(N)} to vanish, 
\be\label{cond:xi}
\frac{\phi_c}\mpl=\sqrt{\frac32}\ln\left(1+\frac{m^2}{3M^2\xi}\right).
\ee
For simplicity, we require that this critical value $\phi_c$ is larger than $\phi_*$, so that $\chi$ will stay
 around $\chi=0$ at the end of the first stage, and will drive the following small-field 
inflation by $\chi$ later. The condition that $\phi_*\lesssim\phi_c$ gives us a constraint on $\xi$,
\be\label{constraint:xi}
\xi\lesssim\frac{m^2}{3M^2\left(F_*(\mu)-1\right)}.
\ee
From (\ref{F*fit}), for $\mu\sim\mathcal{O}(1)$, we have $F_*(\mu)\sim\mathcal{O}(1)$. Hence, under the assumption that $m\ll M$,
this 
implies a very small $\xi$ which we will discuss in more details later.

We can integrate \eqref{chisl(N)} to get the slow-roll solution for $\chi$, 
which gives
\bea\nonumber
\frac{\chi}{\chi_*}&\simeq&1+3\xi(F-F_*)+\frac{2\left(\frac mM\right)^2+\frac43\xi}{\sqrt{45/\mu^2-1}}\left(\text{arctan}\frac{9F-7}{2\sqrt{45/\mu^2-1}}-\text{arctan}\frac{9F_*-7}{2\sqrt{45/\mu^2-1}}\right)\\\label{sol:chi<}
&&+\left(\frac{12\xi}{\mu^2}-\frac{m^2}{M^2}\right)\ln\frac{F-1-4/\mu^2}{F_*-1-4/\mu^2}-\frac23\xi\ln\frac{9F^2-14F+5+20/\mu^2}{9F_*^2-14F_*+5+20/\mu^2}+\cdots,
\eea
where the dots are terms of higher order in $\xi$ and $(m/M)^2$. 
We have already set the condition \eqref{constraint:xi}, which means 
$\chi$ already starts rolling down its potential towards the large-field 
direction at $\chi_*$. However as $\xi$ and $(m/M)^2$ are both very small 
(we will see that they are of order slow-roll parameters in the following), 
we can neglect the background evolution of $\chi$ at the first stage 
when $\phi$ dominates inflation, so that the endpoint of the first stage depends 
solely on $\phi$. Using the $\delta N$-formalism, 
the curvature perturbation for the first stage\footnote{The first stage is defined as the epoch
when $\phi>\phi_*$, where $\phi_*$ can be approximately given from the relation (\ref{F*fit}).} can be calculated as
\bea\nonumber
\mathcal P_\mathcal{R}&=&N_{,\phi}^2\langle\delta\phi^2\rangle
=\frac{3H^2}{32\pi^2\mpl^2}\left(\frac{1-2F^{-1}+F^{-2}\left(1+4/\mu^2\right)}{F^{-1}+F^{-2}\left(1+4/\mu^2\right)}\right)^2\\\nonumber
&=&\frac{3M^2}{128\pi^2\mpl^4}\frac{\left(1-2F^{-1}+F^{-2}\left(1+4/\mu^2\right)\right)^3}{\left(F^{-1}+F^{-2}\left(1+4/\mu^2\right)\right)^2},\\\label{P<(V)}
&=&\frac{V_0}{24\pi^2\mpl^4}\left(\frac{3}{16}\mu^2\right)\frac{\left(1-2F^{-1}+F^{-2}\left(1+4/\mu^2\right)\right)^3}{\left(F^{-1}+F^{-2}\left(1+4/\mu^2\right)\right)^2},
\eea
where the r.h.s. is evaluated at horizon crossing when $N_k=\ln(k/H_k)$, 
and $F\equiv\exp\left(\sqrt{\frac23}\frac\phi\mpl\right)$. 
The slow-roll parameters at this stage are
\bea\label{epsilon1}
\epsilon_H^{(1)}&=&\frac43\frac{\left(F-\left(1+4/\mu^2\right)\right)^2}{\left(F^2-2F+\left(1+4/\mu^2\right)\right)^2},\\\label{eta1}
\eta_H^{(1)}&=&\frac83\frac{F\left(F^2-2F\left(1+4/\mu^2\right)+\left(1+4/\mu^2\right)\right)}{\left(F^2-2F+\left(1+4/\mu^2\right)\right)^2}
\eea
At leading order we can take the logarithmic
 approximation \eqref{phiapprox} that $F=(4/3)(N-N_*)+F_*$, which gives 
the simple relation of $\epsilon$ and $\eta$ depending only 
on $N-N_*$ (and weakly on $\mu$ too). 

After $\phi$ passes the point $\phi_*$, the energy density of the universe
is dominated by the $\chi$-field, so that the relation \eqref{phihidden}
no longer holds. We expect that at this stage,  $\phi$-field will roll down the 
effective potential and get trapped in its local minimum. If its effective mass is small, 
the trapping process is overdamped. 
On the other hand, for a large effective mass, we can expect $\phi$
to undergo damped oscillations around its locally determined minimum
$\phi_g$, or the vacuum expectation value (VEV) determined by $\chi$.

\begin{figure*}\label{fig:motion3d}
  \centering
  \includegraphics[width=.50\textwidth]{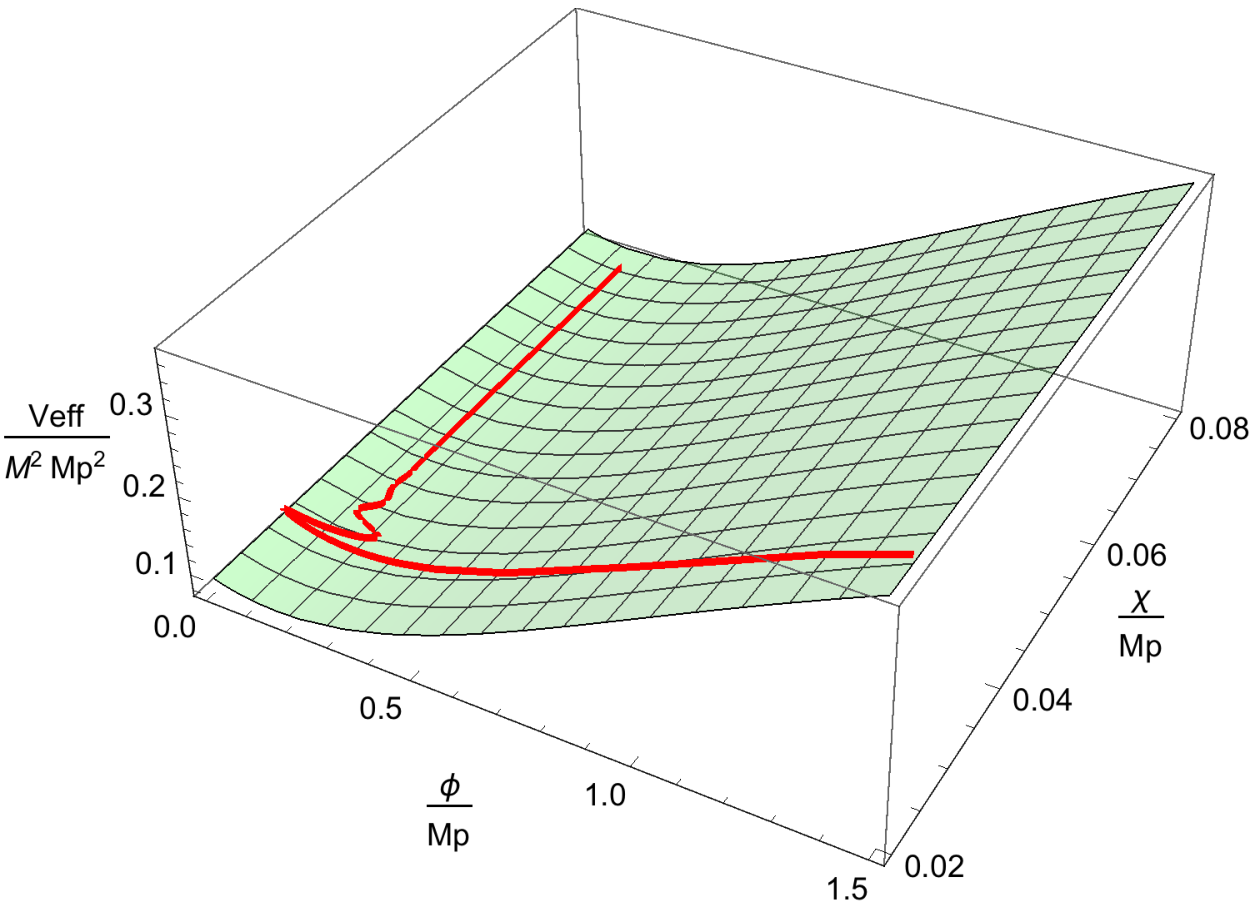}\hfill
   \includegraphics[width=.45\textwidth]{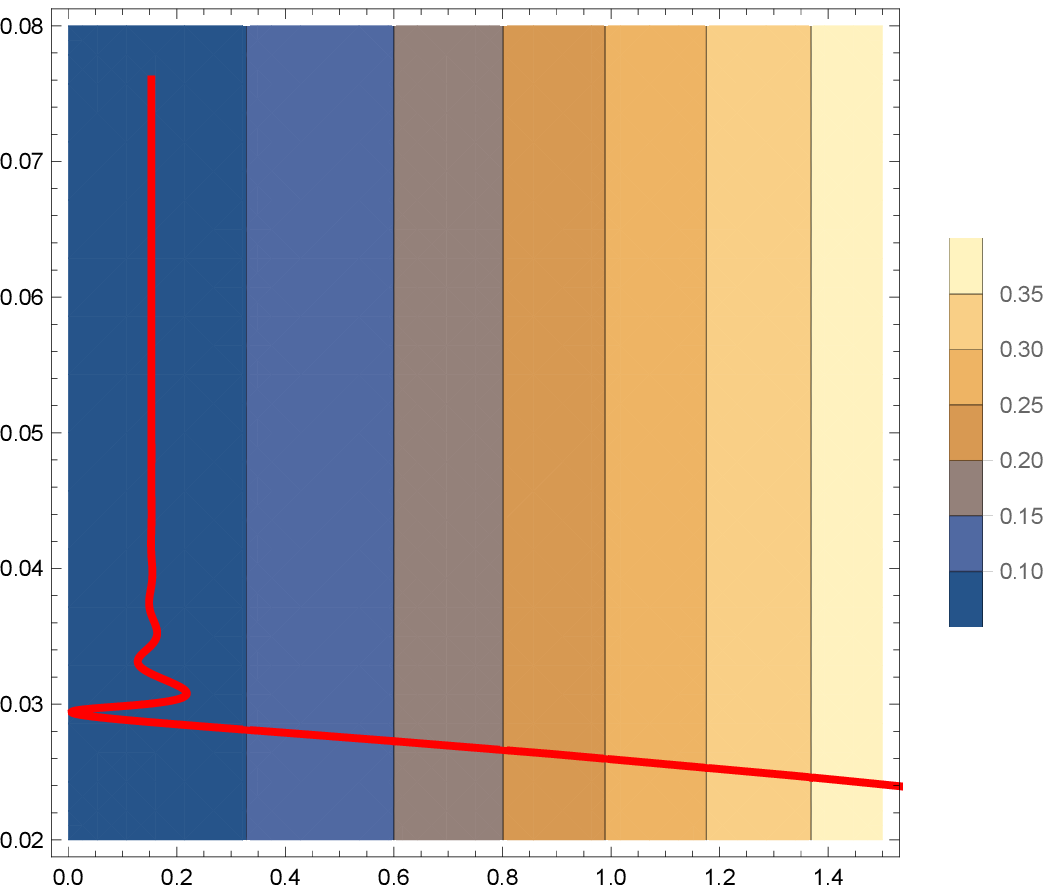}
  \caption{The numerical solutions of $\phi$ and $\chi$ moving on the effective potential $V_\text{eff}$ (left) and in the field space (right), where we have chosen the parameters as $\mu^2=30$, $M/m=20$ and $\xi=0.001$. The initial conditions are taken as $\phi(N_0)=5\mpl$, $\chi(N_0)=0.005\mpl$, $\phi'(N_0)=0.001\mpl$, $\chi'(N_0)=5\times10^{-5}\mpl$, where $N_0=60$, counted backwards from the end of inflation where $N_e=5$. }
\end{figure*}

In \cite{Tolley:2009fg}, the case that $\phi$ is trapped in its local minimum
 is studied,
 while an effective DBI-like action for $\chi$ emerges after 
integrating out $\phi$. This leads to the effective field theory method in the 
literature~\cite{EFT}. Here we use a similar method, but take into account
the oscillations as perturbations to its VEV, for which the 
initial conditions are set at the moment of transition $N_*$.
Namely, the initial amplitude of the oscillation around the local minimum point $\phi_g$,
is given by $\Delta\phi_*=\phi(N_*)-\phi_g(N_*)$. 
$\phi$ will then oscillate around $\phi_g$ with decaying amplitude, 
while the frequency is determined by its effective mass. The numerical solution for the equations of motion of $\phi$ and $\chi$ is depicted in Fig.\ref{fig:motion3d}. As can be seen from this figure, in the second stage, $\phi$-field rolls down the effective potential, then gets trapped in its local minimum point $\phi_g$ and begins to oscillate around it. We will show our analytical result for the oscillation period later.

Let us first calculate $\phi_g$, i.e. 
the background trajectory in the absence of the oscillations.
 The effective potential for $\phi$ is
\be\label{Veff}
V_\text{eff}(\phi)=e^{-\sqrt{\frac23}\frac\phi\mpl}X+\frac34M^2\mpl^2\left[1-e^{-\sqrt{\frac23}\frac\phi\mpl}\left(1-\xi\frac{\chi^2}{\mpl^2}\right)\right]^2+e^{-2\sqrt{\frac23}\frac\phi\mpl}V(\chi),
\ee
where $X=-(1/2)g^{\mu\nu}\partial_\mu\chi\partial_\nu\chi$, and
$\chi$ is assumed to be sufficiently slowly varying.
Assuming that the effective mass of the $\phi$ field is large compared to 
the Hubble parameter, we can immediately find the
minimum of the effective potential,
\be\label{phig}
\frac{\phi_g}{\mpl}=\sqrt{\frac32}\ln\frac{\displaystyle \left(1-\xi\frac{\chi^2}{\mpl^2}\right)^2+\frac{4V(\chi)}{3M^2\mpl^2}}{\displaystyle 1-\xi\frac{\chi^2}{\mpl^2}+\frac{2X}{3M^2\mpl^2}}.
\ee
Substitute this back into \eqref{SE}, we obtain the effective action 
for the $\chi$ field,
\be\label{Seff}
S_g=\int d^4x\sqrt{-g}\left\{\frac{\mpl^2}{2}R+\frac{\displaystyle X\left(1-\xi\frac{\chi^2}{\mpl^2}\right)+\frac{X^2}{3M^2\mpl^2}-V(\chi)}{\displaystyle \left(1-\xi\frac{\chi^2}{\mpl^2}\right)^2+\frac{4V(\chi)}{3M^2\mpl^2}}\right\}.
\ee
This is in the form of an effective action, $f(\chi)X+g(\chi)X^2+V(\chi)$, 
a spectial type of $k$-inflation~\cite{kinf}. 
However, as we focus on the case of slowly rolling $\chi$,
we do not take into account the non-trivial form of the kinetic term.
We leave the study of $k$-inflation for future work. 

From the effective action \eqref{Seff}, under the slow-roll approximation, the Friedmann equation is deduced as follows
\be\label{Friedmann<}
3\mpl^2H_g^2=\frac{\displaystyle V(\chi)}
{\displaystyle \left(1-\xi\frac{\chi^2}{\mpl^2}\right)^2+\frac{4V(\chi)}{3M^2\mpl^2}}.
\ee
We note that $H_g$ in \eqref{Friedmann<} coincides with
the Hubble parameter for the first stage \eqref{H>} 
in the limit $F\rightarrow F(\phi_g)$ where $\phi_g$ is given by \eqref{phig}, 
although \eqref{H>} is only valid for $F>F_*>F(\phi_g)$.
The equation of motion for $\chi$ is
\be\label{eom:chi>}
\left[\left(1-\xi\frac{\chi^2}{\mpl^2}\right)^2+\frac{4V}{3M^2\mpl^2}\right]\left(\ddot\chi+3H\dot\chi\right)+\left(1-\xi\frac{\chi^2}{\mpl^2}\right)^2V,_\chi+\frac{4\xi\chi V}{\mpl^2}\left(1-\xi\frac{\chi^2}{\mpl^2}\right)=0.
\ee
Under the slow-roll condition for $\chi$, we can neglect the 
$\ddot\chi$-term. After changing the time variable to the $e$-folding 
number $N$ defined in \eqref{def:N}, the equation above reduces to the following form
\be\label{eom:chi2}
\frac{d\chi}{dN}=\mpl^2\left(1-\xi\frac{\chi^2}{\mpl^2}\right)^2\frac{V'(\chi)}{V(\chi)}+4\xi\chi\left(1-\xi\frac{\chi^2}{\mpl^2}\right).
\ee

We know from \eqref{constraint:xi} that $\xi\lesssim\frac{m^2}{2M^2}$, which, 
together with $\chi\ll\mpl$ in small field inflation tells us that we can neglect 
the terms involving $\xi$ in \eqref{eom:chi2}. Hence, for a concave
 potential $V=V_0-\frac12m^2\chi^2$, the slow-roll solution for 
the second stage is
\be\label{sol:chi>}
\chi=\chi_*\exp\left[-\frac{m^2\mpl^2}{V_0}\left(N-N_*\right)\right]=\chi_*e^{-\frac{\eta_H^{(2)}}{2}(N-N_*)},
\ee
where $\eta_H^{(2)}$ is the slow-roll parameter in the second stage defined 
from \eqref{eom:chi>}:
\bea\label{def:epsilon<}
\epsilon_H^{(2)}&\equiv&-\frac{\dot H_g}{H_g^2}=\frac1{2\mpl^2}\left(\frac{d\chi}{dN}\right)^2\frac{1}{1+4/\mu^2}=\frac{m^4\mpl^2}{2V_0^2}\frac{\chi^2}{1+4/\mu^2}=\frac{\eta_H^{(2)}{}^2}{8}\frac{(\chi/\mpl)^2}{1+4/\mu^2}.\\\label{def:eta<}
\eta_H^{(2)}&\equiv&\frac{\dot\epsilon_H^{(2)}}{H_g\epsilon_H^{(2)}}=-\frac{d\ln\epsilon_H^{(2)}}{dN}=\frac{2m^2\mpl^2}{V_0}.
\eea
From \eqref{def:eta<}, it is obvious that $\epsilon_H^{(2)}$ is much smaller 
than $\eta_H^{(2)}$ as it is of order $\eta_H^{(2)}{}^2$. Therefore, 
the spectral index
\be
n_s-1=-2\epsilon_H^{(2)}-\eta_H^{(2)}
\ee
is mainly determined by the $\eta_H^{(2)}$ term, 
which gives $\eta_H^{(2)}\sim0.04$ for the current observation data 
if the pivot scale leaves the horizon during the
$\chi$-dominated stage~\cite{Ade:2015lrj}. 
This indicates $\epsilon_H^{(2)}<0.0002$, which is consistent with the 
observational bound $\epsilon_H^{(2)}\lesssim0.006$ given by the constraint 
from the tensor-to-scalar ratio $r=16\epsilon_H^{(2)}\lesssim0.1$. 
Note that if $k_*$ is small and not in the range for CMB and LSS observations, 
then we do not have any strict constraints for 
$\epsilon^{(2)}_H$ and $\eta^{(2)}_H$ in principle. Another quantity 
to be estimate is the upper limit of $\xi$ given in \eqref{constraint:xi}. 
Comparing \eqref{def:eta<} to \eqref{constraint:xi}, we obtain
\be\label{temp1}
\xi\lesssim\frac{\eta_H^{(2)}}{\mu^2\left(F_*(\mu)-1\right)}.
\ee
So we derive that $\xi\lesssim\eta_H^{(2)}/(\mu)^{2}\sim0.01$. This justifies 
why we may neglect the $\xi$-terms in \eqref{eom:chi2}.

As we discussed before, $\phi$ is not equal to $\phi_g$ at the beginning. It rolls down to it from the plateau, and start
to oscillate around $\phi_g$. We set $\phi=\phi_g+\Delta\phi$, 
where the amplitude of the oscillation $\Delta\phi$ is small enough so that it can be treated
as a perturbation. The initial amplitude is given by
$\Delta\phi(N_*)=\phi_*-\phi_g$, where $\phi_*$ is the value
at which the first slow-roll stage ends. Then we have $S=S_g+\Delta S$, where
\be
\Delta S=\int d^4x\sqrt{-g}\left\{-\frac12(\partial\Delta\phi)^2-\frac34M^2\mpl^2\left(1-e^{-\sqrt{\frac23}\frac{\Delta\phi}{\mpl}}\right)^2
\frac{\displaystyle \left(1+\frac{2X}{3M^2\mpl^2}\right)^2}{\displaystyle 1+\frac{4V}{3M^2\mpl^2}}\right\}.
\ee
The e.o.m. for $\Delta\phi$ is then
\be\label{eom(t)}
\ddot{\Delta\phi}+3H\dot{\Delta\phi}+\sqrt{\frac32}M^2\mpl\left(1-e^{-\sqrt{\frac23}\frac{\Delta\phi}{\mpl}}\right)e^{-\sqrt{\frac23}\frac{\Delta\phi}{\mpl}}\frac{\displaystyle \left(1+\frac{2X}{3M^2\mpl^2}\right)^2}{\displaystyle 1+\frac{4V}{3M^2\mpl^2}}=0.
\ee
Changing the time variable to the $e$-folding number $N$, neglecting
the slow-roll suppressed terms, \eqref{eom(t)} reduces to the following form
\be\label{eom:Dphi(n)}
\frac{d^2}{dN^2}\frac{\Delta\phi}{\mpl}-3\frac{d}{dN}\frac{\Delta\phi}{\mpl}+\sqrt{\frac32}\mu^2\left(1-e^{-\sqrt{\frac23}\frac{\Delta\phi}{\mpl}}\right)e^{-\sqrt{\frac23}\frac{\Delta\phi}{\mpl}}
=0.
\ee
This is a second-order differential equation for nonlinear oscillation. 
The analytical result can only be found when the initial amplitude
\be\label{Deltaphicontinuous}
\Delta\phi(N_*)=\phi_*-\phi_g=\mpl\sqrt{\frac32}~\ln\frac{F_*(\mu)}{1+4/\mu^2},
\ee
is sufficiently smaller than $\mpl$. Hence, we can 
treat $\Delta\phi$ as a small quantity, and approximate 
\eqref{eom:Dphi(n)} by the corresponding linear equation:
\be\label{eom:Dphi(n)linear}
\frac{d^2}{dN^2}\frac{\Delta\phi}{\mpl}-3\frac{d}{dN}\frac{\Delta\phi}{\mpl}+\mu^2\frac{\Delta\phi}{\mpl}=0.
\ee
The continuity of the first derivative of the $\phi$ field is that $\Delta\phi'(N_*)$ equals to $\phi'(N_*)$ in the first stage given in \eqref{phidot*}. 
Together with the matching conditions \eqref{Deltaphicontinuous} and \eqref{phidot*} 
we obtain
\be\label{Deltaphi(chi)}
\frac{\Delta\phi}\mpl=e^{\frac32(N-N_*)}\sqrt{\frac32}\ln\frac{F_*(\mu)}{1+\frac4{\mu^2}}\sqrt{1+\Upsilon^2}\cos\left[\sqrt{\mu^2-\frac94}(N-N_*)+\text{arctan}\Upsilon\right],
\ee
where $F_*(\mu)$ is given in \eqref{F*fit} and
\bea\label{def:Upsilon}
\Upsilon&=&\frac1{\sqrt{\mu^2-9/4}}\left[\frac32-\frac43\frac{F_*-1-4/\mu^2}{F_*^2-2F_*+1+4/\mu^2}\left(\ln\frac{F_*}{1+4/\mu^2}\right)^{-1}\right].
\eea
From the ``frequency'' term in \eqref{Deltaphi(chi)}, in order for 
the oscillations to happen, $\mu>3/2$ is required, 
which gives the lower bound of $\mu$.  This analytical solution \eqref{Deltaphi(chi)} can be compared with the numerical solutions depicted in Fig.\ref{fig:motion3d}.  

\section{Perturbation and the Features on the Power Spectrum}\label{sec:pert}
We use the $\delta N$ formalism~\cite{Sasaki:1995aw} to calculate the power spectrum 
of the curvature perturbation at the second stage when $\chi$ dominates 
the energy density. Here we have
\be\label{RdN}
\mathcal{R}=\delta N=N_{,\chi}\delta\chi+N_{,\phi}\delta\phi+\cdots.
\ee
$N_{,\phi}$ is much smaller than $N_{,\chi}$, since $\phi$ only vibrates around its fixed trajectory $\phi_g$. Besides, the contribution to the curvature perturbation from $\langle\delta\phi\rangle$ is suppressed by $e^{-\pi\mu/2}$ at the horizon crossing and is negligible~\cite{Mukhanov:2007zz}. Therefore in this second stage, the curvature perturbation is mainly contributed by the fluctuations of $\chi$. However, there are corrections coming from the oscillating background of $\phi$ via the Hubble parameters, slow-roll parameters, and, most importantly, the $\dot\phi$-$\dot\chi$ coupling. 
Under the slow-roll approximation, \eqref{eom:phi} reduces to the following form
\be
3H\left(1-\frac13\sqrt{\frac23}\frac{\dot\phi}{H\mpl}\right)\dot\chi+e^{-\sqrt{\frac23}\frac\phi\mpl}V'(\chi)=0.
\ee
Changing the variable to $N$, and using the Friedmann equation \eqref{Friedmann<}, we obtain
\be
\left(1+\frac13\sqrt{\frac23}\frac1\mpl\frac{d\phi}{dN}\right)\frac{d\chi}{dN}=\mpl^2e^{-\sqrt{\frac23}\frac{\Delta\phi}\mpl}\frac{V'(\chi)}{V(\chi)}.
\ee
If the dynamical coupling is small compared to the background evolution of $\chi$, we can treat the $d\phi/dN$ term as well as its derivative as a small quantity and find the linearized equation:
\be
\frac{d\ln\chi}{dN}=-\frac{\eta_H^{(2)}}{2}\left(1-\sqrt{\frac23}\frac{\Delta\phi}{\mpl}-\frac13\sqrt{\frac23}\frac1\mpl\frac{d\phi}{dN}\right),
\ee
where $\eta_H^{(2)}$ is the second slow-roll parameter defined by \eqref{def:eta<}. This equation can be easily solved, which gives the inverse $N(\chi)$ relation as
\be
N_0-N=\frac2{\eta_H^{(2)}}\ln\frac\chi{\chi_0}+\sqrt{\frac23}\int^N_{N_0}\frac{\Delta\phi}{\mpl}dN+\frac13\sqrt{\frac23}\frac{\phi-\phi_0}{\mpl},
\ee
where $N_0$ is a reference $e$-folding number and $\phi_0=\phi(N_0)$. Then, we take the derivative of $N$ with respect to $\chi$, and obtain the linearized result
\be
\frac{\partial N}{\partial\chi}=-\frac{2}{\eta_H^{(2)}\chi}\left(1
+\sqrt{\frac23}\frac{\Delta\phi}{\mpl}+\frac13\sqrt{\frac23}\frac{d}{dN}\frac{\Delta\phi}{\mpl}\right).
\ee
From the action \eqref{SE}, it is obvious that the canonical field is not $\chi$, but $\hat\chi=F^{-1/2}\chi$.
This gives a normalization factor in the amplitude of the quantum flactuation of $\chi$:
\be
\langle\delta\chi\delta\chi\rangle=F\langle\delta\hat\chi\delta\hat\chi\rangle=e^{\sqrt{\frac23}\frac\phi\mpl}\left(\frac{H}{2\pi}\right)^2\approx\left(1+\frac4{\mu^2}\right)\left(1+\frac12\sqrt{\frac23}\frac{\Delta\phi}{\mpl}\right)^2\left(\frac{H}{2\pi}\right)^2.
\ee
Therefore the power spectrum of the curvature perturbation on the superhorizon scale is
\bea\nonumber
\mathcal P_\mathcal{R}&=&\left(\frac{\partial N}{\partial\chi}\right)^2\langle\delta\chi\delta\chi\rangle,\\\nonumber
&=&\mathcal P^{(0)}_\mathcal{R}\left\{1
-\frac13e^{\frac32(N-N_*)}\ln\frac{F_*}{1+\frac4{\mu^2}}\sqrt{1+\Upsilon^2}\bigg[\omega\sin\left(\omega(N-N_*)+\tan^{-1}\Upsilon\right)\right.\\
&&-\left.\left.6\cos\left(\omega(N-N_*)+\tan^{-1}\Upsilon\right)\right]\right\}^2,
\eea
where $\mathcal P_\mathcal{R}^{(0)}$ is the power spectrum from the background energy density:
\be
\mathcal P_\mathcal{R}^{(0)}\equiv\frac{H^2}{8\pi\epsilon_H^{(2)}\mpl^2}\approx\frac{V_0}{24\pi^2\mpl^2}\left(\frac{\mpl}{\chi_*}\right)^2\frac{8}{\eta_H^{(2)}{}^2}e^{\frac{\eta_H^{(2)}}{2}(N-N_*)},
\ee
with $\epsilon_H^{(2)}$ defined in \eqref{def:epsilon<}, $\eta_H^{(2)}$ defined in \eqref{def:eta<}, and $H$ given in \eqref{Friedmann<}. Note that all the quantities are calculated at the moment of horizon-crossing when $k=H_ka_k$, which gives
\begin{flalign}\nn
&&\frac{\mathcal P_\mathcal{R}}{\mathcal P^{(0)}_\mathcal{R}}=\left\{1+\frac13\left(\frac{k_*}{k}\right)^{\frac32}\ln\frac{F_*}{1+\frac4{\mu^2}}\sqrt{1+\Upsilon^2}\left[\omega\sin\left(\omega\ln\frac{k}{k_*}+\tan^{-1}\Upsilon\right)\right.\right.\\
&& \left.\left.+6\cos\left(\omega\ln\frac{k}{k_*}+\tan^{-1}\Upsilon\right)\right]\right\}^2,\label{P(k)}
\end{flalign}
where $F_*$ is given in \eqref{F*fit}, $\Upsilon$ is defined in \eqref{def:Upsilon}, and the frequency of the oscillation is $\omega=\sqrt{\mu^2-9/4}$.

\subsection{Large Scale Anomaly in CMB anisotropies}\label{sec:cmb}

\begin{figure*}
  \centering
  \includegraphics[width=.46\textwidth]{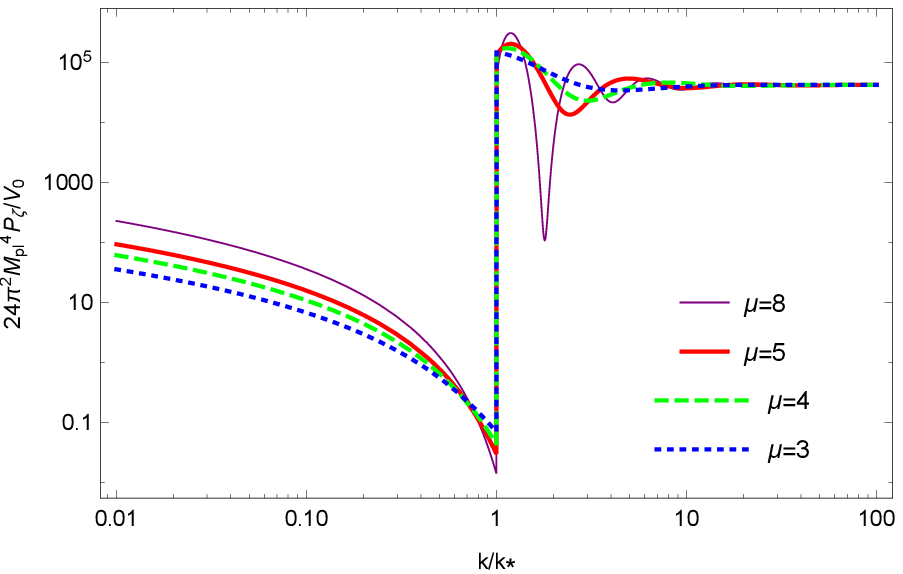}\hfill
  \includegraphics[width=.48\textwidth]{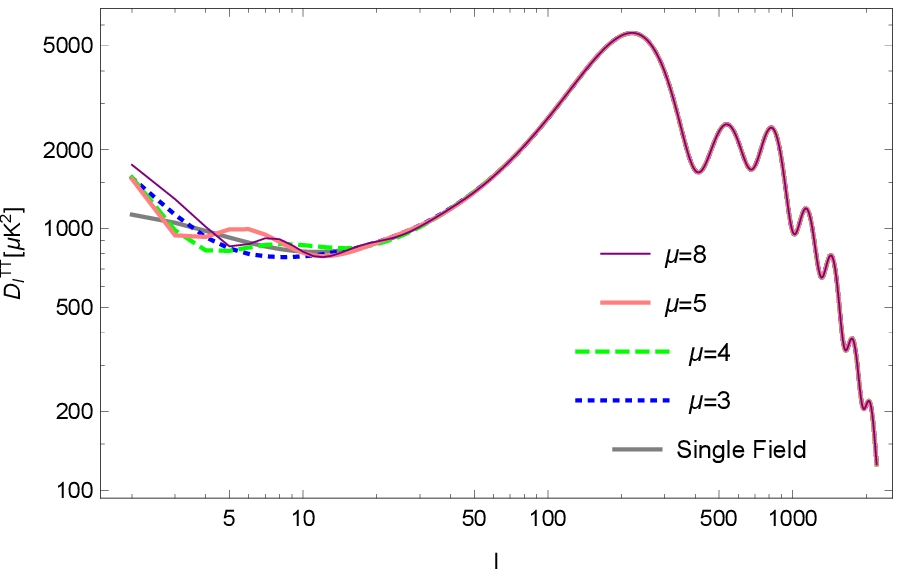}
  \caption{Left: The oscillation features in the power spectrum for the curvature perturbation with different values of $\mu$. The result is robust in other parameters as we can see in \eqref{P(k)}. A huge enhancement of $(\eta_H^{(2)})^{-2}$ has appeared around the junction of the two stages $k_*$. In this plot we choose $\mpl/(\eta^{(2)}_H\chi_*)=200$. Right: CMB temperature anisotropy when the fluctuation with wavenumber $k_*$ is right outside our observable universe today, which is set to $k_*=10^{-4}$~Mpc${}^{-1}$.
  }\label{fig:P}
\end{figure*}

We depict the power spectrum and the corresponding CMB angular spectrum
in Fig.~\ref{fig:P}. The key characteristic of the power spectrum is that there is a huge enhancement at around the scale of $k_*$, which is of order $(\eta_H^{(2)})^{-2}(\mpl/\chi_*)^{2}$. Then the spectrum begins oscillating, with the amplitude of the corrections of order 10 to $10^2$, depending sensitively on $\mu$. The correction is so large that $K_*$ is only possible to appear on the scales larger, or smaller than the range detected by CMB or LSS. For the former possibility, we draw on the right panel of Fig.~\ref{fig:P} the CMB temperature anisotropy in case when $1/k_*$ is right beyond the largest observable scale, say $k_*\sim10^{-4}~\text{Mpc}^{-1}$. We see that $C_{TT}$ oscillates at around $l<20$, which is possible to explain the low-$l$ anomaly on the CMB anisotropies.

We can see that this is different from the oscillation amplitude generated by a massive field with a power-law potential during inflation~\cite{standardclock, Huang:2016quc}.  There is no slow-roll suppression for the amplitude in our model because of the $\dot\phi$-$\dot\chi$ coupling, which originates from the non-trivial metric induced by the $R^2$-gravity. Such a coupling which is absence in Einstein theory will contribute an extra friction term of order $M$ to the equation of motion for $\chi$, and amplifies the correction in the final power spectrum. 

\subsection{Primordial Black Holes as Dark Matter}\label{sec:pbh}

\begin{figure*}
\centering
   \includegraphics[width=.55\textwidth]{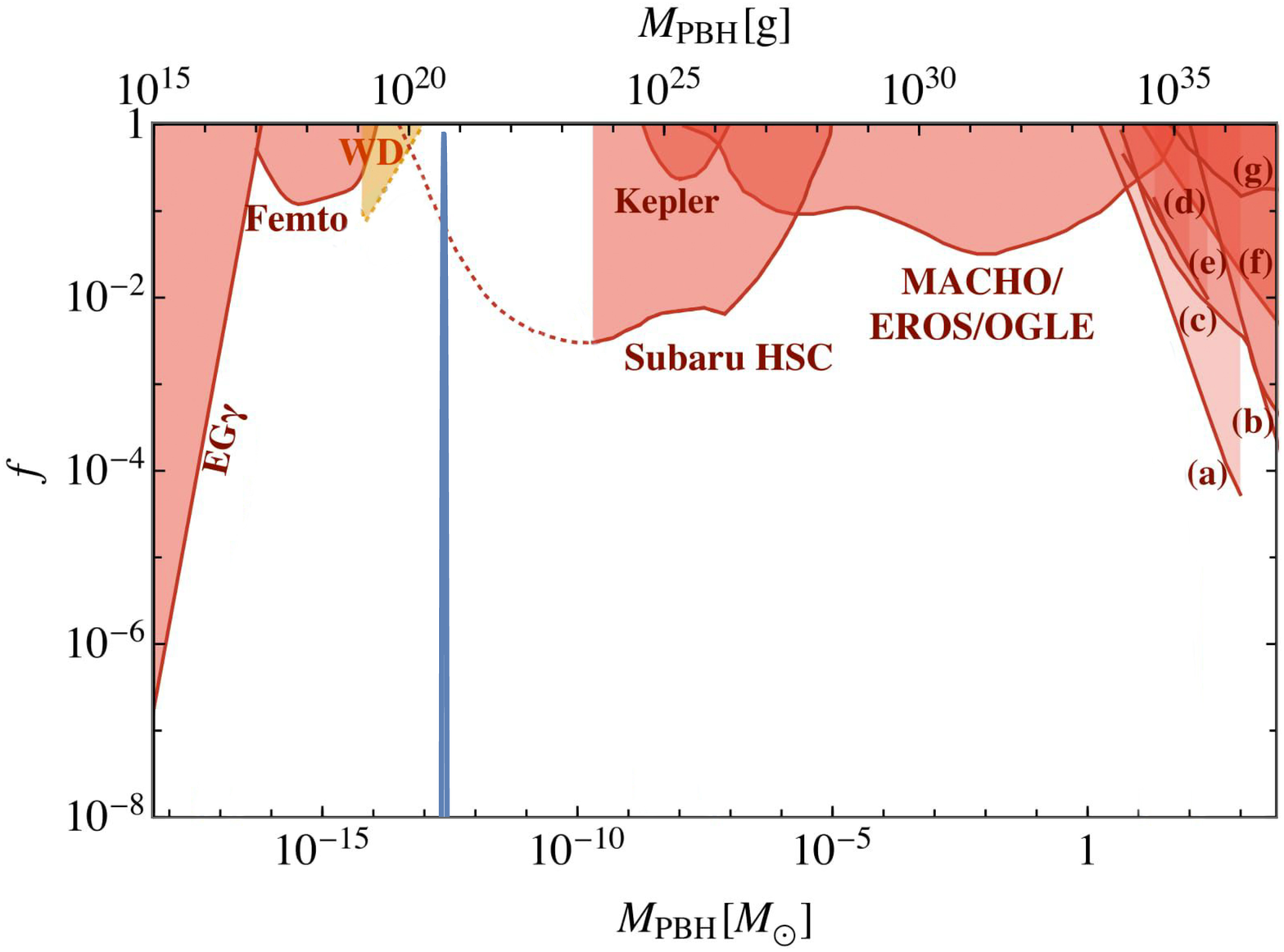}\hfill
  \includegraphics[width=.45\textwidth]{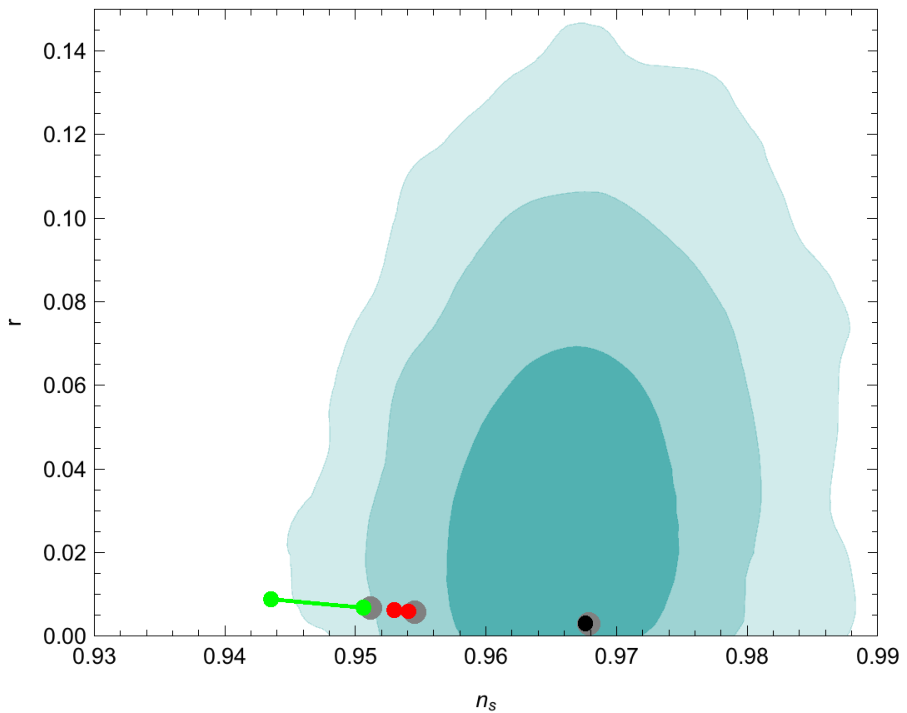}
  \caption{Left: The constraints on the primordial black hole mass density contrast to the dark matter density $f(M)=\Omega_\text{PBH}/\Omega_\text{CDM}$ from \cite{Inomata:2017bwi}, with $f(M)$ derived in our model (blue sharp peak). We choose $\mu=8$, $\mathcal{P_R}^{(0)}=8.9\times10^{-4}$, $\delta_\text{c}=0.45$, and $\Delta N_1=37$, which corresponds to the peak of $f=0.73$ at $2.8\times10^{20}~\text{g}$. We see that it is monochromatic since its width is $\Delta\ln M\sim0.1$. The red dotted curve shows the weakened constraint of Subaru HSC because of the wave effect~\cite{Inomata:2017okj,Inomata:2017bwi,Takada:2017}. Right: The tensor-to-scalar ratio ($r$) vesus spectral tilt ($n_s$), with dependence on the $e$-folding numbers endured in the first stage of inflation. When $e$-folding number of the first stage decreases, the spectral tilt will also decrease. We draw especially the interesting cases for $N_1=42\sim41$(red dots), $39\sim34$ (green dots) and $22\sim19$ (left beyond the frame), 
which correspond to the mass of PBH $10^{16}\sim10^{17}~\text g$, 
$10^{19}\sim10^{23}~\text g$, and $1\sim10^{3}M_\odot$, respectively. 
The black dot is our model with $\Delta N_1=60$, while the bigger gray dots are $\Delta N_1=60,~42,~39,~22$ (from right to left) in the original $R^2$-inflation. The shaded contours at 1, 2, and 3-$\sigma$ confidence level from BICEP2/Keck Array and Planck joint analysis~\cite{Ade:2015tva,Jiang:2017nou}.}\label{fig:rns}
\end{figure*}

Another possible application is when the transition happens late in inflation, which means the large scale observational data are mainly contributed by the first stage of inflation. This will make the enhancement of power spectrum at a very small scale beyond the CMB or LSS observations, where there is basically only very loose constraint~\cite{Bringmann:2011ut}. However, in this case primordial black holes may be generated when the scales with wavenumber around $k_*$ re-enter the horizon, if the power spectrum of the density perturbation has exceeded some threshold $\delta_\text{c}\sim0.45$~\cite{Hawking:1971ei,Carr:1974nx,Meszaros:1974tb,Carr:1975qj}. The primordial black holes can be the candidates of a part or all of the dark matter~\cite{pbhdm,GarciaBellido:1996qt,Green:2004wb}, which have attracted much attention especially after the detection of gravitational waves~\cite{Abbott:2016blz}\cite{ligodm}.  If we assume this re-entry happens during the radiation dominated epoch, the mass of the primordial black hole generated for a fluctuation which leaves the horizon $\Delta N_1$ efolds after the pivot scale horizon-exit is
\be\label{M(k)}
\Delta N_1\simeq41-\frac12\ln\left(\frac{M_\text{PBH}}{10^{16}~\text{g}}\right)+\frac12\ln\mu.
\ee
See Appendix \ref{app:pbh} for the derivation and discussion of \eqref{M(k)}. Setting the mass spectrum to peak at the possible windows in the PBH mass fraction,  we can connect the $e$-folding number of the first stage with the position of the peak on the PBH mass. The height of the peak is determined by both the enhancement of the power spectrum and the first oscillation peak of it, which can be tuned by the parameters and made large enough to generate adequate PBHs as dark matter. The width for the peak can be estimated by $\Delta \ln M\sim\Delta N\sim1/\mu$, which is of order $0.1$ in our model, thus it is reasonable to use the monochromatic assumption. In these cases, we can go back to \eqref{epsilon1} and \eqref{eta1} to depict the tensor-to-scalar ratio and the spectral tilt for different $N_1$, inspired by possible masses of the peaks for the PBH fraction as dark matter. Especially, recent study on the wave effect implies there is no lensing magnification on the Subaru Hyper Supreme-Cam (HSC) constraint, which opens the ``window'' for the PBH mass fraction as dark matter at around $10^{19}~\text g$ to $10^{23}~\text g$~\cite{Inomata:2017okj,Inomata:2017bwi,Takada:2017}. This provides possibility that PBHs can be the candidate for all the dark matter at this broad mass range,  of which the predictions of $r$ and $n_s$ are inside the 3-$\sigma$ contour of Bicep/Keck+Planck joint data in our model as can be seen in Fig.~\ref{fig:rns}. 

\section{Conclusion}\label{sec:con}
In this paper, we studied a model with $R^2$-gravity plus a scalar field 
$\chi$ non-minimally coupled to $R$. This is partly motivated by going beyond 
the $R^2$-inflation, and partly by our interest in the effect
of a heavy field with a Starobinsky-like potential. 
We focused on the case when the scalaron $\phi$ turns into a heavy field
during inflation, and solved the evolution of the system including 
its perturbation. In this case, inflation will be split into two different 
stages dominated by the scalaron $\phi$ and the other field $\chi$, respectively. 

The main result is that there appears an enhancement of the curvature
perturbation power spectrum as well as an oscillatory feature. 
Unlike the usual case of a minimally coupled heavy field, 
these features are not suppressed by slow-roll parameters.
This implies it is difficult to accommodate these features on the CMB scale 
since they are too large to be consistent with observation.
 However, it is possible that the transition happens on scales
beyond the current Hubble scale, and only their ``tails'' 
affect the CMB scale. In this case, they may explain the observed
low-$\ell$ anomalies. 

Another case is when the transition happens on very small scales.
In this case, the enhancement in the power spectrum may lead
to the formation of primordial black holes which may constitute
a large fraction of or even the whole dark matter observed today.
An interesting outcome is that due to a narrow
enhancement feature in the power spectrum, there appears
a sharp peak in the PBH mass spectrum. That is,
PBH masses will be nearly monochromatic.
We found that for PBHs with mass $\sim 10^{22}$g, which 
can account for the whole dark matter according to the
recent analysis~\cite{Inomata:2017okj,Inomata:2017bwi,Takada:2017},
 the predicted $n_s$ is at the edge of the 3$\sigma$ confidence level contour 
of Keck/Bicep+Planck data, 
 while the higher mass range, 
for instance for PBHs of $10\sim10^2$ solar masses to account for
those detected by LIGO~\cite{Abbott:2016blz,ligodm}, 
can not be realized in our model because it can not give
enough number of $e$-folds for the first stage of inflation.

Although the predicted $n_s$ is still within the 3$\sigma$ contour, 
our model with PBHs as dark matter may be excluded as soon as the 
contour shrinks slightly by improvements in the data quality. 
It is therefore useful to consider 
modifications of the model to make it more compatible with the CMB observations. 
This may be realized in models which predict $n_s$ larger than 
the Planck observed value at 60 $e$-foldings. Then the best-fitted point
in the $(n_s,r)$ plane can be realized for smaller numbers
of $e$-folds for the first stage, which leads to larger PBH masses.

\begin{figure*}
\centering
   \includegraphics[width=.6\textwidth]{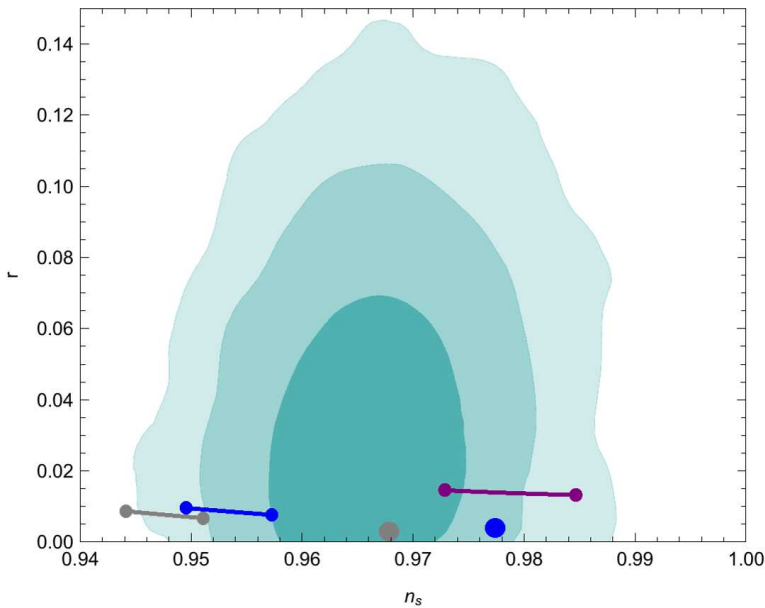}
  \caption{The predictions from a corrected $f(R)$ gravity defined in \eqref{def:R3}. We only depict $N_1=34$ to 39 which corresponds to the interesting PBH mass range $M_\text{PBH}=10^{19}\text g$ to $10^{23}~\text g$, for $\lambda=0$~(Starobinsky model, gray line and dots), $-2\times10^{-5}$~(blue), and $-5\times10^{-5}$~(purple). The larger dots are the reference point for $\Delta N_1=60$.}\label{fig:R3}
\end{figure*}

An interesting possible modification is to consider 
a small $R^3$ correction to the Starobinsky model,
\be\label{def:R3}
f(R)=R+\frac{R^2}{6M^2}+\lambda\frac{R^3}{3M^4}.
\ee
For a small and negative $\lambda$, it makes the potential slightly
shallower and predicts larger $n_s$ without changing
the tensor-to-scalar ratio much. 
We depict in Fig.~\ref{fig:R3} the corresponding predictions 
for $\lambda=0$, $-2\times10^{-5}$, and $-5\times10^{-5}$. 
We see that for an appropriate choice of the cubic term,
 the range for $e$-folding numbers that correponds to the rencent 
discovered window for PBH as dark matter fits the observation well.

In this paper we set an upper bound for the mass parameter $\mu$ of our model
in order not to give rise to a temporary halt of inflation between the two stages. 
There is no fundamental reason, however, for avoiding such a case.
We did not discuss this possibility just for simplicity.
We would like to leave it for future work. 
Also, in this paper we focused on the case where both stages are 
driven by a flat potential with the standard kinetic term, apart from
the non-trivial field space metric. 
It may be interesting to extend our analysis to 
the case when the kinetic term is highly non-trivial, or more generically 
to Horndeski gravity and beyond-Horndeski gravity~\cite{galileon}.

\vspace{5mm}
\noindent {\bf Acknowledgments}

We thank Shinji Tsujikawa and Jun-ichi Yokoyama for useful discussions.  
This work was supported in part by the MEXT KAKENHI No. 15H05888
and 15K21733.
SP is supported by the MEXT KAKENHI No. 15H05888. 
YLZ is supported by the NSFC grant No. 11605228, 11673025, 11720101004. 
QGH is supported by grants from NSFC (grant NO. 11335012, 11575271, 11690021,11647601), 
Top-Notch Young Talents Program of China, and partly supported by Key Research 
Program of Frontier Sciences, CAS.

\appendix
\renewcommand{\thefigure}{A}
\setcounter{figure}{0}
\section{Endpoint of Inflation with Multiple Stages}\label{app:efold}
In this appendix we count the $e$-folding number from the moment when 
a given comoving wavelength exits the horizon to when it re-enters. 
This method was developed in \cite{Liddle:2003as}, and has some recent 
developments \cite{efold}. Here we will recast it with consideration of 
slow-roll corrections. 

We suppose that the inflation consists of two different stages, each of which 
is slow-roll inflation, with slow-roll parameters $\epsilon_1(N)$ and $\epsilon_2(N)$. 
We count the $e$-folding number forward in time from the initial time $a_i$ 
when the comoving scale corresponding to the current Hubble scale $H_0$ 
exits the horizon. 
The Hubble parameter as a function of the $e$-folding number $N$ is expressed as
\bea\label{def:H1}
H_1(N)&=&H_i\exp\left(-\int^N_{N_i}\epsilon_1(N)dN\right),\\
\label{def:H2}
H_2(N)&=&H_i\exp\left(-\int^{N_*}_{N_i}\epsilon_1(N)dN
-\int^{N}_{N_*}\epsilon_2(N)dN\right)
\eea
where $N_i$ is the $e$-folding number at the moment when $H=H_i$, 
and $N_*$ is the transition time between the two stages. 
From the above we find that at the end of inflation when $N=N_e$, 
the Hubble parameter is
\be
H_e=H_2(N_e)=
H_i\exp\left(-\int^{N_*}_{N_i}\epsilon_1(N)dN
-\int^{N_e}_{N_*}\epsilon_2(N)dN\right).
\label{def:He}
\ee
After inflation, it is followed by a (quadratic) oscillation stage where 
the effective equation of state is the same as nonrelativistic matter. 
Let us suppose the oscillation lasts for $\Delta N_\text{osi}$ $e$-foldings 
after which the universe is thermalized and is dominated by radiation, until 
$z=10^4$ when matter becomes dominant.

Following the brief history above, we divided the given comoving 
wavenumber $k$ by the current Hubble scale,
\be
\frac{k}{a_0H_0}
=\frac{a_k}{a_*}\frac{a_*}{a_e}\frac{a_e}{a_{r}}\frac{a_r}{a_\text{0}}\frac{H_k}{H_0},
\ee
where $a_r$ is the scale factor at the end of reheating. See Fig.~\ref{fig:efold}
for the relations between various epochs and the corresponding Hubble scales.
Using the $e$-folding numbers defined above, we can write this as
\be\label{main}
\ln\frac{k}{a_0H_0}=-\Delta N_1-\Delta N_2-\Delta N_\text{osi}
+\ln\frac{a_r}{a_\text{0}}+\ln\frac{H_k}{H_0}.
\ee
$\Delta N_1$ and $\Delta N_2$ are the $e$-folding numbers of the 
first and second stages of inflation, respectively. 
We have defined $\Delta N_\text{osi}\equiv\ln(a_r/a_f)$ as the 
$e$-folding number of the oscillation stage after inflation.
The scale factor at the moment when the universe is thermalized, $a_r$, 
can be calculated by the entropy conservation till now:
\be
g_rT_r^3a_r^3=\left(2+\frac4{11}g_0^\nu\right)T_0^3a_0^3,
\ee
where $g_r$ is the number of degrees of freedom at the moment when 
the universe is thermalized, $g_0^\nu$ is the current degrees of freedom 
of neutrinos, and the factor $4/11$ comes from the temperature difference 
between neutrinos and photons after decoupling of the weak interaction. 
For 3 generations of neutrinos, we have $g_0^\nu=21/4$, which gives
\be\label{ar/a0}
\ln\frac{a_r}{a_0}=\ln\frac{T_0}{\rho_r^{1/4}}
+\ln\left(\frac{43}{11g_r}\right)^{1/3}\left(\frac{\pi^2g_r}{30}\right)^{1/4},
\ee
where we have used the Stephan-Boltzmann law
\be
\rho_r=g_r\frac{\pi^2}{30}T_r^4
\ee
to express the temperature of the universe when reheating completes 
in terms of the energy density at that moment. The $\rho_r$ in \eqref{ar/a0} can 
be expressed as
\be\label{rhof}
\ln\frac{\rho_e}{\rho_r}=-3\ln\frac{a_e}{a_r}=3\Delta N_\text{osi}.
\ee
Substitute \eqref{rhof} into \eqref{ar/a0}, then back into \eqref{main}, 
we have
\be\label{main2}
\Delta N_1+\Delta N_2+\frac{\Delta N_\text{osi}}4-\ln\frac{H_k}{\rho_e^{1/4}}
=\ln\frac{T_0}{H_0}-\ln\frac{k}{a_0H_0}
+\ln\left(\frac{43}{11g_r}\right)^{1/3}\left(\frac{\pi^2g_r}{30}\right)^{1/4}.
\ee
To estimate the last term on the l.h.s., we use the Friedmann equation 
at $N_e$, $\rho_f=3H_f^2\mpl^2$, and also the expression for the tensor 
power spectrum
\be
\mathcal P_h=\frac{2}{\pi^2}\left(\frac{H_k}{\mpl}\right)^2=r\mathcal{P_R}.
\ee
Then we can express $\ln(H_k/\rho_e^{1/4})$ as
\be
\ln\frac{H_k}{\rho_e^{1/4}}=-\ln3^{1/4}+\frac12\ln\frac{\pi}{\sqrt2}\left(r\mathcal{P_R}\right)^{1/2}+\ln\frac{H_k^{1/2}}{H_e^{1/2}}.
\ee
The last term can be estimated by the expression of $H(N)$ we obtained in \eqref{def:H2}, which gives
\be
\ln\frac{H_k^{1/2}}{H_e^{1/2}}=\frac12\int_{N_k}^{N_*}\epsilon_1(N)dN+\frac12\int_{N_*}^{N_e}\epsilon_2(N)dN.
\ee
Therefore we have the result,
\bea
&&\Delta N_1+\Delta N_2+\frac{\Delta N_\text{osi}}{4}-\frac12\int_{N_k}^{N_*}\epsilon_1(N)dN-\frac12\int_{N_*}^{N_e}\epsilon_2(N)dN\\\nonumber
&=&\ln\frac{T_0}{H_0}-\ln\frac{k}{a_0H_0}+\ln\frac{43^{1/3}\pi}{11^{1/3}180^{1/4}}-\frac1{12}\ln g_r+\frac14\ln\left(r\mathcal{P_R}\right).\\\label{main3}
&\approx&60-\frac1{12}\ln\frac{g_r}{10^2}+\frac14\ln\frac{r}{0.01}\equiv N_\text{tot}.
\eea
In the last line we use the Planck 2015 observational data $T_0=2.73\text{K}$, $H_0=67.8\text{km/s/Mpc}$, $\mathcal{P_R}=2.21\times10^{-9}$ as well as the pivot scale $k=0.002\text{Mpc}^{-1}$, and calculate the pure number $N_\text{tot}$. The dependence on $g_r$ and $r$ are rather weak, and we can neglect them in the following discussions on premordial black holes production. The schematic diagram of \label{main3} can be seen in Fig.\ref{fig:efold}.

\begin{figure}

\centering

\includegraphics[width=.7\textwidth]{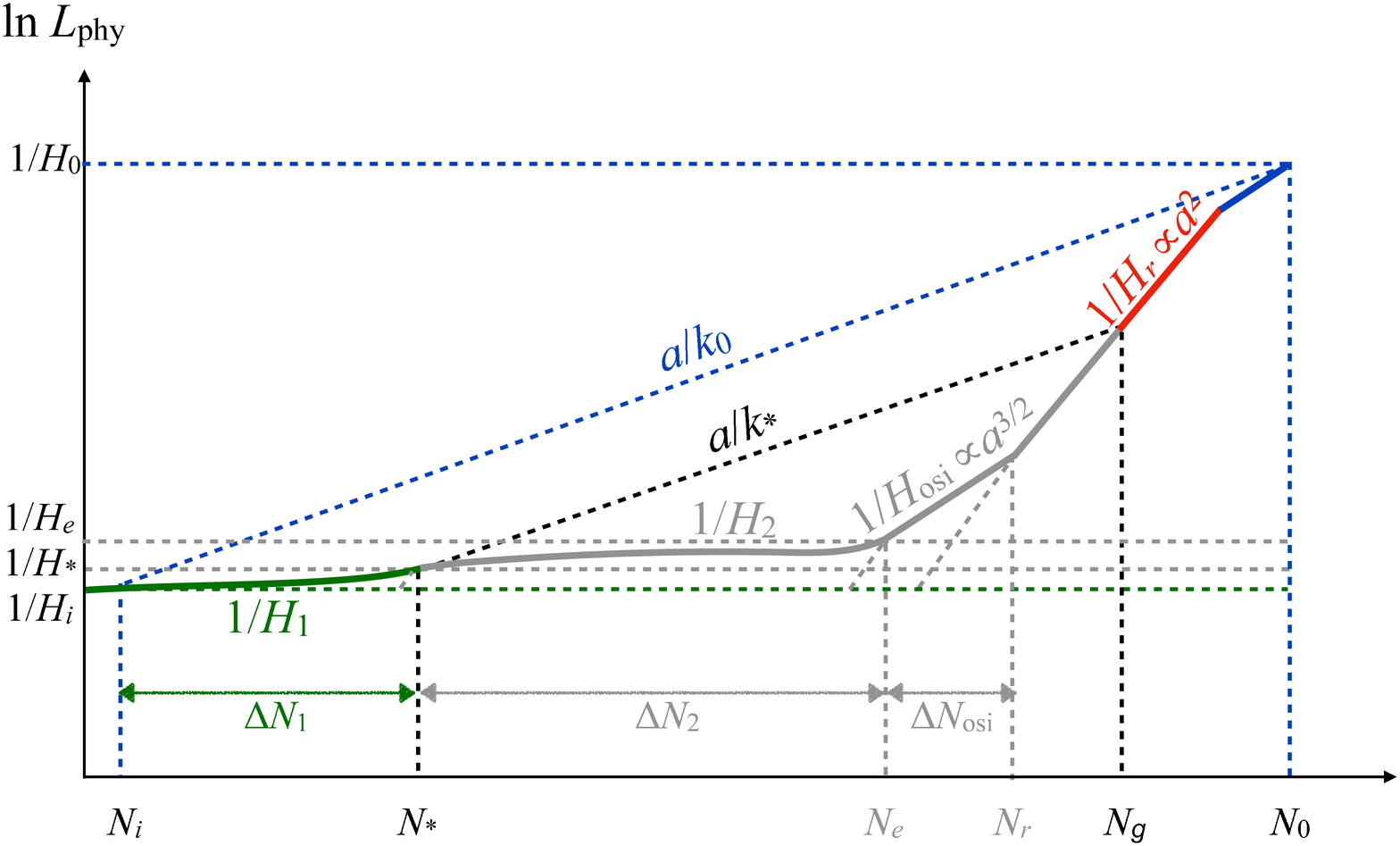}

\caption{Schematic diagram of $H(N)^{-1}$ in the expansion history of our universe, taking into account the two different stages of inflation. $N_i$ is some reference point we start to count $e$-folding numbers, $N_*$ is the transition of the two stages, $N_e$ is the end of inflation, $N_r$ is the moment when the universe gets completely thermalized, and $N_g$ is the moment when the wavenumber which exits the horizon at $N_*$ re-enters the horizon. It is also the time for primordial black holes formation if there is a peak in the power spectrum at $k_*$. The Green curve to the left is $H_1(N)^{-1}$ defined in \eqref{def:H1}, while the red line is $H^{-1}$ at radiation-dominated epoch, thus proportional to $a^2$. The gray curves are the evolution of $H^{-1}$ from $N_*$ to $N_g$, which is irrelevant to the PBH mass if $k_*$ re-enters the horizon at radiation-dominated epoch.}

\end{figure}\label{fig:efold}

\section{Mass Formula for Primordial Black Holes}
\label{app:pbh}
In this appendix we derive a formula for the mass of primordial black holes 
produced at a given moment in the radiation dominated era in terms of
a relevant number of $e$-folds during inflation.
 
The idea of PBHs~\cite{Hawking:1971ei} as dark matter was studied in 
1970s~\cite{Carr:1974nx,Meszaros:1974tb,Carr:1975qj}, and recent observational 
constraints, epecially the detection of gravitational waves from
binary BH mergers by LIGO/VIRGO, have inspired some more discussions~\cite{ligodm}. 
In \cite{GarciaBellido:1996qt},  the PBH mass is estimated as
\be\label{app:mass1}
M_\text{PBH}\simeq\frac{\mpl^2}{H_*}e^{2N_c},
\ee
where 
$H_*$ is the Hubble parameter when the $k_*$-mode leaves the horizon during inflation, 
with $k_*=a_*H_*$ being the wavenumber at which the primordial power spectrum peaks, 
and $N_c$ is the $e$-folding number from the horizon exit to the end of inflation. 
This relation is derived under the assumption that all the energy density 
has been converted to radiation right after inflation.
Below we extend it to the general case. Namely, we derive a formula
that is independent of reheating process after inflation.

Let us label the moment of horizon re-entry for the mode $k_*$ corresponding 
to the peak in the power spectrum as $a_g$. The PBH mass at formation is 
proportional to the total mass inside the horizon at $a_g$, 
i.e. $M_\text{PBH}=\gamma M_\text{horizon}$. Here $\gamma$ is a numerical 
factor depending on the collapse mechanism, and a reasonable estimation could 
be made for $\gamma\sim0.2$ to 0.4~\cite{Carr:1975qj,Green:2004wb}. 
Then the PBH mass at formation can be estimated by
\be\label{app:mhorizon}
M_\text{PBH}=\gamma\frac\pi6\rho_g\left(\frac{a_g}k\right)^3,
\ee
where $\rho_g$ can be connected to $\rho_r$, thus to $\rho_\text{osi}$, by
\be
\rho_g=\rho_r\left(\frac{a_r}{a_g}\right)^4
=\rho_e\left(\frac{a_e}{a_r}\right)^3\left(\frac{a_r}{a_g}\right)^4.
\ee
At the end of inflation, from the Friedmann equation, $\rho_e=3\mpl^2H_e^2$, 
we can then use \eqref{def:H2} to get
\be
\rho_e=3\mpl^2H_e^2=3\mpl^2H_*^2\exp\left(-2\int^e_*\epsilon(N)dN\right).
\ee

 Another relation can be found by calculating the ratio of the Hubble 
radius from $a_*$ to $a_g$ by two different methods: one is to evaluates $H_g$
 from the evolution law of \eqref{def:H2} during inflation, $H\propto a^{-3/2}$ 
during the oscillatory stage,
 and $H\propto a^{-2}$ during the radiation dominated universe, 
while the other is to simply use the relation $a_*H_*=a_gH_g$. 

Equating these two ratios, we have
\be
-\int^e_*\epsilon(N)dN-\frac32\Delta N_\text{osi}-2\ln\frac{a_g}{a_r}
=-\Delta N_2-\Delta N_\text{osi}-\ln\frac{a_g}{a_r},
\ee
which gives
\be
\ln\frac{a_g}{a_r}=\Delta N_2-\int^e_*\epsilon(N)dN-\frac{\Delta N_\text{osi}}{2}.
\ee
Therefore we can substitute these relations back into \eqref{app:mhorizon} to get
\be\label{app:mass2}
M_\text{PBH}=\gamma
\frac{\mpl^2}{H_*}\exp\left[2\left(
\Delta N_2+\frac{\Delta N_\text{osi}}{4}-\frac12\int^e_*\epsilon(N)dN\right)\right].
\ee
This is an extension of \eqref{app:mass1} which takes into account 
both the small time variation of the Hubble parameter
durin inflation and the existence of the oscillation stage 
after inflation. The correction of $\Delta N_\text{osi}$ depends on the details, 
especially the temerature of reheating/preheating, thus can 
be of order $\mathcal{O}(10)$ when the reheating temperature is 
low~\cite{efold}. The corrections from the slow-roll parameter may also 
be non-negligible. Since the slow-roll parameter does not 
have to be much smaller than 1 on small scales, 
a value of $\epsilon\lesssim1$ may change the PBH mass significantly. 

Both $\Delta N_2$ and $\Delta N_\text{osi}$ depend on the physics on small scales 
and are difficult to estimate. However, it is interesting to note
that \eqref{app:mass2} can be written in the form that depends only 
on the physics on large scales. To see this, we notice that the 
combination in the exponent can be expressed in terms of
similar quantities on large scales by the relation \eqref{main3},
 which gives a simpler version of $\eqref{app:mass2}$,
\be\label{app:mass3}
M_\text{PBH}\simeq\frac{\mpl^2}{H_*}\exp\left[2\left(N_\text{tot}-\Delta N_1+\frac12\int^{\Delta N_1}_0\epsilon(N)dN\right)\right].
\ee
where both the total number of $e$-folds $N_\text{tot}$ and the integral 
are counted from the same initial moment $N=N_i$. 

Actually the fact that the formula is independent of the small scale physics 
is very general.
It is completely independent of the evolutionary behavior of the
universe during the period from the epoch $N=N_*$ to the PBH formation epoch
as long as the universe has become radiation dominated by that time.
This can be seen from Fig.~\ref{fig:efold}.
For the pivot scale $0.002~\text{Mpc}^{-1}$, we have $N_\text{tot}\sim60$ as 
shown in Appendix \ref{app:efold}. 
The only fact to be kept in mind is that it is derived under the assumption
of radiation dominance when the wavenumber $k_*$ re-enters the horizon.
If $k_*$ re-enters in the earlier oscillation stage before the universe is 
thermalized, we should turn to a modified 
formula $M_\text{PBH}=\gamma(\mpl^2/H_*)e^{3\Delta N_2}$, 
where $\Delta N_2<\Delta N_\text{osi}/2$. In this case, the details of 
reheating inevitably will enter, in addition to the fact that
the coefficient $\gamma$ may not be uniquely defined~\cite{GarciaBellido:1996qt,pbhrh}.

In our model, the PBH mass depends only on the $e$-folding number and 
the slow-roll parameter at the first stage.
For simplicity we can borrow the result from pure $R^2$-inflation,
\be
\epsilon_1\simeq\frac{12}{(4N_1+3)^2}\;,
\quad
\eta_1\simeq\frac{16 (2 N_1+3)}{(4 N_1+3)^2}\;,
\ee
to obtain
\be
M_\text{PBH}\simeq\gamma\frac{\mpl^2}{H_k}
e^{2\left(N_\text{tot}-\Delta N_1+\frac{1/2}{1+3/(4\Delta N_1)}\right)}
\ee
Conversely for a given PBH mass, the corresponding $e$-folding number
of the first stage is expressed as
\be\label{main5}
\Delta N_1\simeq41
-\frac12\ln\left(\frac{M_\text{PBH}}{10^{16}~\text{g}}\right)+\frac12\ln\mu.
\ee
The PBHs lighter than $10^{16}~\text g$ has already evaporated by today. 
For a given $\mu$, we can use this relation to estimate the $e$-folding number 
from the horizon exit of the pivot scale to the end of the first stage 
as a function of the PBH mass. For instance, for $\mu=8$, in the possible window 
of the PBH mass as dark matter, we find $M_{PBH}=10^{16}\sim10^{17}$g, 
$10^{19}\sim10^{23}$g, and $1\sim10^3\,M_\odot$,
correspond to $\Delta N_1=41\sim42$, $34\sim39$, and $19\sim22$, respectively.

\end{document}